\documentclass[aps,prl,twocolumn,showpacs,superscriptaddress]{revtex4-2}

\usepackage{graphicx}
\usepackage{dcolumn}   
\usepackage{bm}        
\usepackage{amssymb}   
\usepackage{amsmath}
\usepackage{url}
\usepackage{layout}
\usepackage{color}
\usepackage{colortbl}
\usepackage{epsfig}
\usepackage{tabularx}
\usepackage{mathrsfs}

\usepackage{appendix}
\usepackage{xcolor}
\usepackage[thinlines]{easytable}
\usepackage{makecell}
\usepackage{tikz,pgf}
\usepackage{booktabs}
\usepackage{float}
\usepackage{hyperref}
\hypersetup{colorlinks,citecolor=blue}

\begin{document}
\title{Pairing mechanism and superconductivity in 1313 phase La$_3$Ni$_2$O$_7$}

\author{Cui-Qun Chen}
\thanks{These two authors contributed equally to this work.}
\affiliation{Guangdong Provincial Key Laboratory of Magnetoelectric Physics and Devices, State Key Laboratory of Optoelectronic Materials and Technologies, Institute of Neutron Science and Technology, School of Physics, Sun Yat-sen University, Guangzhou 510275, China}

\author{Ming Zhang}
\thanks{These two authors contributed equally to this work.}
\affiliation{Zhejiang Key Laboratory of Quantum State Control and Optical Field Manipulation,
Department of Physics, Zhejiang Sci-Tech University, 310018 Hangzhou, China}

\author{Fan Yang}
\email{yangfan\_blg@bit.edu.cn}
\affiliation{School of Physics, Beijing Institute of Technology, Beijing 100081, China}

\author{Dao-Xin Yao}
\email{yaodaox@mail.sysu.edu.cn}
\affiliation{Guangdong Provincial Key Laboratory of Magnetoelectric Physics and Devices, State Key Laboratory of Optoelectronic Materials and Technologies, Institute of Neutron Science and Technology, School of Physics, Sun Yat-sen University, Guangzhou 510275, China}

\begin{abstract}
We systematically investigate the electronic properties and superconducting mechanism of 1313 La$_3$Ni$_2$O$_7$ using density functional theory plus dynamical mean-field theory (DFT+DMFT) and random phase approximation (RPA).
Our DFT+DMFT calculations reveal that the single-layer (SL) subsystem exhibits nearly insulating behavior, with the $d_{z^2}$ orbital showing Mott physics, while the trilayer (TL) subsystem remains metallic. This indicates that SC primarily resides in the TL subsystem, whose Ni-$e_g$ orbitals are found to be hole-doped relative to bulk La$_4$Ni$_3$O$_{10}$. 
Based on DFT+DMFT-derived low-energy Hamiltonian, RPA-based analysis yields an $s^{\pm}$-wave pairing symmetry within the TL subsystem. Importantly, we identify two key factors that contribute to the significant suppression of $T_c$ in 1313 La$_3$Ni$_2$O$_7$ compared to bulk La$_4$Ni$_3$O$_{10}$. First, the hole doping in the TL subsystem leads to a decreased pairing strength. Second, the SL subsystem acts as a bridge connecting adjacent superconducting TL subsystems, thereby forming an S-N-S Josephson junction. The resulting interlayer Josephson coupling governs the phase coherence between TL subsystems and further suppresses the global $T_c$.  Combinedly, our findings suggest that the high-$T_c$ phase in the RP La$_3$Ni$_2$O$_7$ family should be attributed to the 2222 La$_3$Ni$_2$O$_7$ rather than the 1313 La$_3$Ni$_2$O$_7$.

\end{abstract}
\maketitle

{\it Introduction.} Since the discovery of superconductivity (SC) up to 78 K in pressurized La$_3$Ni$_2$O$_7$~\cite{1}, Ruddlesden-Popper (RP) nickelates have emerged as a hot research focus, triggering an intense surge of both theoretical and experimental investigations~\cite{2,3,4,5,6,7,  8,9,10,11,12,13,14,15,16,17,18,19,20,21,22,23}. 
Whether SC can also be realized in hybrid RP nickelates has become an important question of interest~\cite{24, 25,26, 27, 28}. 
Experimentally, pressurized and thin-film 1212 La$_5$Ni$_3$O$_{11}$ exhibits SC up to 64 K and 50 K~\cite{29}. 
However, there is ongoing theoretical debate regarding the pairing mechanism in 1212 La$_5$Ni$_3$O$_{11}$. A random phase approximation (RPA)-based study attributed the SC to the single-layer (SL) subsystem, but the resulting pairing strength is too weak to account for the experimentally observed high $T_c$~\cite{30}. Other works suggest that SC in 1212 La$_5$Ni$_3$O$_{11}$ arises primarily from the bilayer subsystem~\cite{28, 31,32}, with the SL subsystem forming a Josephson coupling structure with the bilayer subsystem, which explains the dome-shaped dependence of $T_c$ observed in experiments~\cite{31}.

\begin{figure}[b]
\centering
\includegraphics[width=0.9\columnwidth]{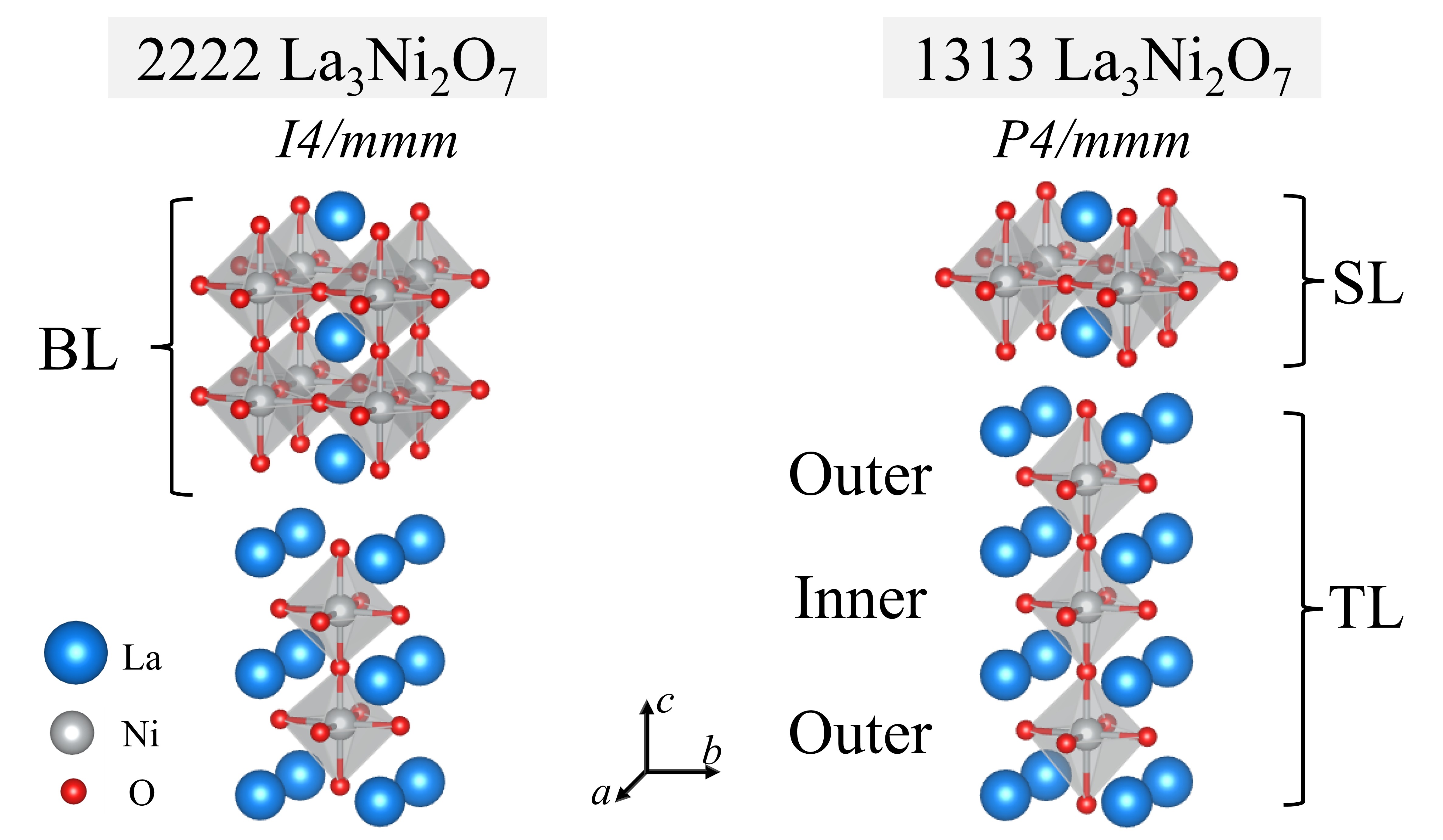}
\caption{Crystal structures of high-pressure phase  of 2222 and  1313 La$_3$Ni$_2$O$_7$. The blue, gray, and red balls represent La, Ni, and O atoms, respectively. }
\label{fig1}
\end{figure}

\begin{figure*}
\noindent \begin{centering}
\includegraphics[width=2\columnwidth,height=2\columnwidth,keepaspectratio]{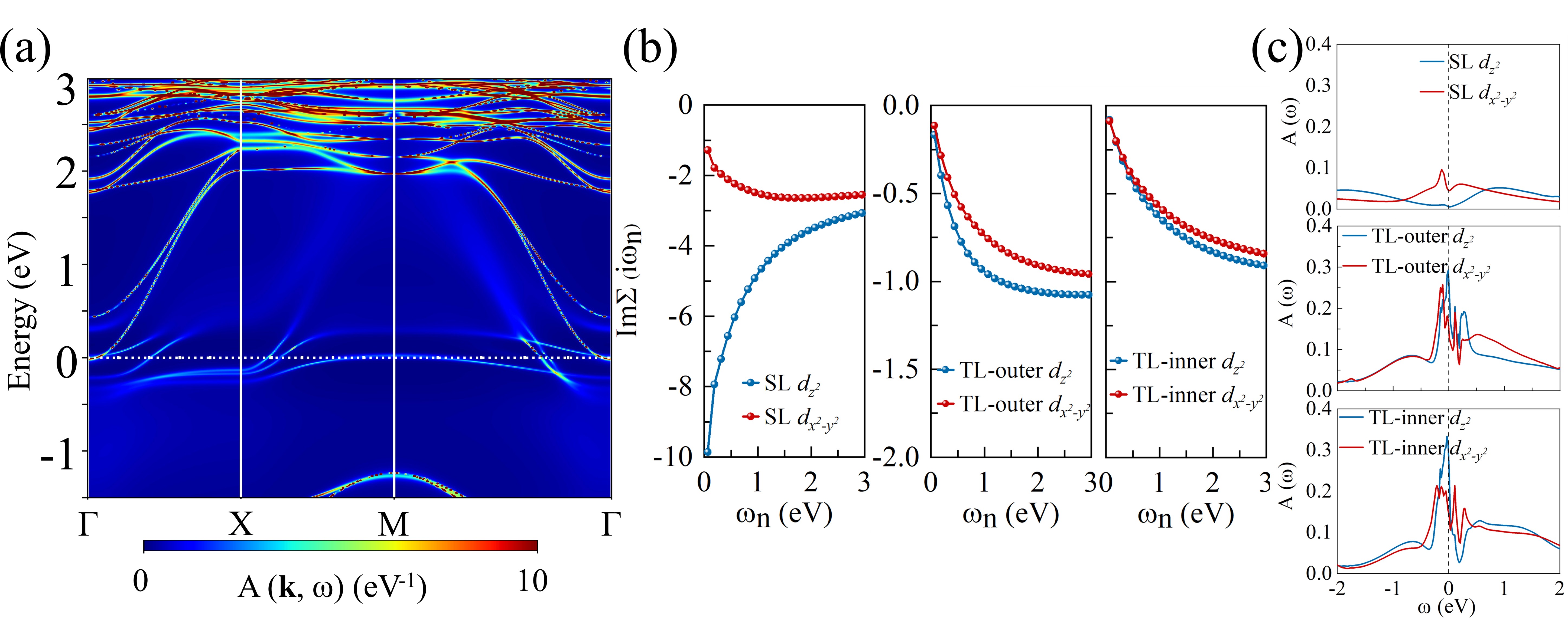}\caption{ DFT+DMFT calculated momentum-resolved spectral functions $A(\mathbf{k},\omega)$ (a), imaginary parts of Matsubara self-energy functions $\mathbf{Im}\sum i\omega_n$ (b) and  orbital-resolved spectral functions $A(\omega)$ (c) at 230 K under 20 GPa. The inner and outer layers are indicated in Fig.\ref{fig1}
  \label{fig2}}
\par\end{centering}
\end{figure*}

Recently, the SC of 1313 phase of La$_3$Ni$_2$O$_7$ has attracted significant attention. A previous experimental study reported filamentary SC up to 80 K in 1313 La$_3$Ni$_2$O$_7$~\cite{26}. However, it was noted that the possibility of high-$T_c$ arising from other phases could not be ruled out. More recently, a new experiment has demonstrated that 1313 La$_3$Ni$_2$O$_7$ exhibits SC with a maximum $T_c$ of 3.6 K above 19 GPa~\cite{33}. 
Moreover, thin-film experiments have revealed that while the 2222 phase of La$_3$Ni$_2$O$_7$ can manifest SC, the 1313 phase does not~\cite{34,35}, casting further doubt on the intrinsic SC of 1313 La$_3$Ni$_2$O$_7$. Theoretically, studies on the nature of SC in 1313 La$_3$Ni$_2$O$_7$ have yielded conflicting conclusions. An earlier RPA-based study proposed that the SL subsystem hosts a leading $d_{x^2-y^2}$ pairing state, with the nonsuperconducting trilayer (TL) subsystem in-between~\cite{36}. In contrast, density functional theory plus dynamical mean-field theory (DFT+DMFT) investigations have revealed that the SL subsystem exhibits Mott insulating behavior, rendering it an unlikely host for SC~\cite{37,38,39}. Therefore, a comprehensive understanding of the electronic and superconducting properties of 1313 La$_3$Ni$_2$O$_7$ becomes an urgent and crucial issue in the field.

In this work, we systematically study the electronic properties using DFT+DMFT calculations, followed by an RPA based analysis to investigate the superconducting mechanism in hybrid 1313 La$_3$Ni$_2$O$_7$.
DMFT results reveal a layer-dependent correlation effect, where the SL subsystem manifests as a nearly insulating bad metal,
while the TL subsystem remains metallic and features hole-doped Ni-$e_g$ orbitals relative to bulk La$_4$Ni$_3$O$_{10}$.
Using DFT+DMFT derived effective model, RPA analysis identifies an $s^{\pm}$-wave pairing symmetry in the TL subsystem of 1313 La$_3$Ni$_2$O$_7$. The pairing strength within the TL subsystem shows a decreasing trend upon hole doping, implying a lower $T_c$ compared to bulk La$_4$Ni$_3$O$_{10}$. Additionally, the SL and TL subsystems form an S-N-S Josephson junction, where the SL subsystem acts as a weak link, suppressing phase coherence and leading to a significantly reduced global $T_c$. Our findings clarify the genuine high-$T_c$ phase in RP La$_3$Ni$_2$O$_7$ series and highlight the important difference between hybrid-phase and pure-phase nickelate superconductors.

\label{sec:electronic}

\begin{table}[t]
\renewcommand{\arraystretch}{1.2}
\caption{The DFT+DMFT-calculated Ni-$e_g$ orbital occupations for TL sybsystem of 1313 La$_3$Ni$_2$O$_7$ under 20 GPa and La$_4$Ni$_3$O$_{10}$ under 44.3 GPa \cite{40}. }
\noindent\begin{centering}
\begin{tabular*}{\columnwidth}{@{\extracolsep{\fill}}cccc@{}}
\hline \hline 
 &  $n(d_{z^{2}})$ & $n(d_{x^{2}-y^{2}})$  & $n(\mathrm{total})$   \tabularnewline 
  \hline 
1313 La$_3$Ni$_2$O$_7$ TL-inner & 1.040  &  0.986 &  2.026  \tabularnewline 
1313 La$_3$Ni$_2$O$_7$ TL-outer &  1.032 &  1.055 &  2.087  \tabularnewline 

La$_4$Ni$_3$O$_{10}$-inner  & 1.060  & 1.017  & 2.077    \tabularnewline 
La$_4$Ni$_3$O$_{10}$-outer  & 1.088  & 1.060  & 2.148 \tabularnewline 
\hline \hline 
\end{tabular*}
\label{tab:occ}
\par\end{centering}
\end{table}

{\it Electronic correlation.} Hybrid 1313 La$_3$Ni$_2$O$_7$ features an alternating stacking structure composed of SL La$_2$NiO$_4$ and TL La$_4$Ni$_3$O$_{10}$ subsystems, with a translation of $(\frac{1}{2},\frac{1}{2})$ between SL and TL subsystems, as shown in Fig.\ref{fig1}. 
At ambient pressure, 1313 La$_3$Ni$_2$O$_7$ crystallizes in the orthorhombic $Cmmm$ space group~\cite{33}. By applying pressure,  a structural phase transition occurs at approximately 13 GPa, leading to a
tetragonal phase with $P4/mmm$ space group.

Figure \ref{fig2}(a) presents the charge fully self-consistent DFT+DMFT calculated correlated electronic structure of the 1313 La$_3$Ni$_2$O$_7$ at 20 GPa. The momentum-resolved spectral function $A(\mathbf{k},\omega)$ reveals pronounced band renormalization effects, with the Ni-$e_g$ bandwidth reduced from 4 eV in DFT calculation to approximately 2.5 eV in DMFT calculation.
Around the M point, the bands originating from the TL subsystem become flattened due to renormalization effects, and the bonding band slightly crosses the Fermi level, leading to the $\gamma$ pocket on the Fermi surface (FS).

To further elucidate the characteristics of the bands near the FS, Figure \ref{fig2}(b) and (c) show the orbital-resolved imaginary part of Matsubara self-energy function $\mathrm{Im}\sum (i\omega_n)$ and orbital-resolved spectral functions $A(\omega)$. The self-energy function of the SL subsystem exhibits distinct behaviors between $d_{z^2}$ and $d_{x^2-y^2}$ orbitals at low frequencies. Specifically, the SL-$d_{z^2}$ orbital manifests Mott physics, whereas the SL-$d_{x^2-y^2}$ orbital displays a large intercept at zero-frequency, indicating the  intrinsic strong correlations of the SL subsystem. 
Note that the $d_{x^2-y^2}$ orbital exhibits a tiny  spectral function intensity in the vicinity of the Fermi energy.
However, the spectrum is dispersed throughout the entire momentum space, without manifesting well-defined quasi-particle excitation, as illustrated in Fig.\ref{fig2}(a). 
This behavior is fundamentally different from that of a doped Mott insulator that becomes superconducting upon carrier doping, where coherent quasiparticles typically emerge. Here, despite charge transfer between two subsystems, the SL subsystem remains incoherent. Moreover, ARPES measurements have not detected any Fermi surface from the SL subsystem~\cite{34}. These observations collectively rule out the SL subsystem as a viable host for superconductivity.
Therefore, the SL subsystem behaves as a nearly insulating bad metal, characterized by a gap near the Fermi level and the large imaginary part of self-energy.

\begin{figure}[t]
\centering
\includegraphics[width=0.9\columnwidth]{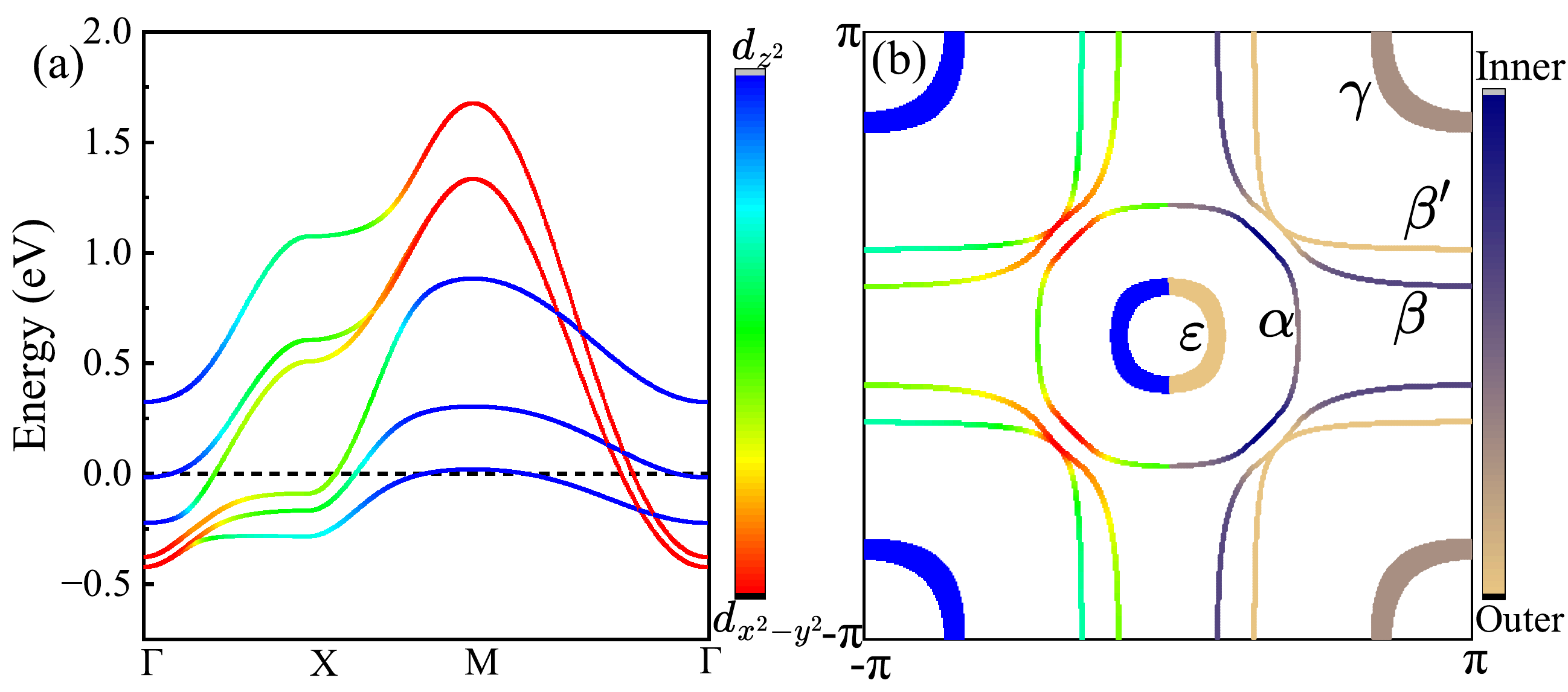}
\caption{Renormalized TL TB band structure (a) and FS (b). The color bars indicate the orbital weight of Ni-$d_{x^{2}-y^{2}}$ and $d_{z^{2}}$, and the relative contributions from Ni atoms in the outer-layer and inner-layer, respectively. }
\label{fig3}
\end{figure}

Regarding the TL subsystem, the self-energy functions reflect that it belongs to a correlated metal.
In particular, the outer layer Ni-$e_g$ orbitals display stronger correlations with a slightly larger intercept at zero frequency, similar to the bulk La$_4$Ni$_3$O$_{10}$~\cite{40}. The spectral function of the $d_{z^2}$ orbital in the TL subsystem exhibits a peak at the Fermi level, corresponding to the two flat bands, bonding and nonbonding bands, at the M point in Fig.\ref{fig2}(a). Collectively, these findings suggest that SC in the 1313 La$_3$Ni$_2$O$_7$ should originate primarily from the TL subsystem, while the SL subsystem is unlikely to host the superconducting carriers.

The local occupations of Ni-$e_g$ orbitals of TL subsystem are listed in Table \ref{tab:occ}, together with the occupations of bulk La$_4$Ni$_3$O$_{10}$ from previous study for comparison~\cite{40}. The DFT+DMFT calculated occupations of the Ni-$e_g$ orbitals exceed their nominal values (1.33 electrons per Ni), implying the $d$-$p$ orbital hybridization. 
Compared to bulk La$_4$Ni$_3$O$_{10}$, the Ni-$e_g$ orbital occupancy in the TL subsystem is reduced by approximately 0.17, indicating that the Ni-$e_g$ orbitals of TL subsystem is hole-doped relative to the bulk La$_4$Ni$_3$O$_{10}$. 
Note that this reduction reflects the actual decrease after fully accounting for $d$-$p$ hybridization, and thus should be distinguished from the electron filling used in the tight-binding (TB) model for SC calculations discussed below.

\label{sec:model}

{\it Renormalized trilayer two-orbital model.} Given that SC is likely to primarily originate from the TL subsystem, we construct a TL two-orbital model to investigate the electronic properties based on the DFT+DMFT results as described in Supplementary Materials~\cite{41}. Taking the Ni-$e_g$ orbitals as basis, the Hamiltonian can be expressed as follows:
\begin{eqnarray}
\label{eq1}
H_0=\sum_{i,\mu,\sigma}\epsilon_{\mu}\hat{c}^{\dagger}_{i\mu\sigma}\hat{c}_{i\mu\sigma}+\sum_{i,j,\mu,\nu,\sigma}t_{ij}^{\mu\nu}(\hat{c}^{\dagger}_{i\mu\sigma}\hat{c}_{j\nu\sigma}+h.c),
\end{eqnarray}
where $i/j$, $\sigma$, and $\mu/\nu$ denote the indices of site, spin, and orbital, respectively. $\epsilon_{\mu}$ represents the on-site energy of the orbital $\mu$. 

\begin{table}[t]
\renewcommand{\arraystretch}{1.5}
\caption{Parameters of the renormalized  TB model for 1313 La$_3$Ni$_2$O$_7$ in unit of eV. $x$ and $z$ label the $d_{x^{2}-y^{2}}$ and $d_{z^{2}}$ orbitals. I and O represent the innerlayer and outerlayer. $\epsilon$ denotes the on-site energy. $t_z^{TL}$ indicates the intertrilayer hopping of $d_{z^{2}}$ orbital.}
\noindent\begin{centering}
\begin{tabular*}{\columnwidth}{@{\extracolsep{\fill}}cccccccc@{}}
\hline \hline 
 $t_{[1,0]}^{x,O}$ & $t_{[1,0]}^{z,O}$ & $t_{[1,1]}^{x,O}$ & $t_{[1,1]}^{z,O}$ & $t_{[2,0]}^{x,O}$ & $t_{[1,0]}^{xz,O}$ & $t_{[2,0]}^{xz,O}$ & $t_{[1,0]}^{x,I}$\tabularnewline 
-0.220 & -0.040 & 0.028 & -0.007 & 0.030 & -0.096   & -0.013 & -0.257\tabularnewline 
\hline 
   $t_{[1,0]}^{z,I}$ & $t_{[1,1]}^{x,I}$ & $t_{[2,0]}^{x,I}$ & $t_{[1,1]}^{z,I}$ & $t_{[1,0]}^{xz,I}$ & $t_{[2,0]}^{xz,I}$ &  $t_{[0,0]}^{z,IO}$ & $t_{[0,0]}^{z,OO}$ \tabularnewline 
 -0.060 & 0.030 & -0.036 & -0.009 & 0.133   & 0.019 & -0.223 & -0.024 \tabularnewline 
\hline 
 $t_{[1,0]}^{xz,IO}$ & $t_{[1,0]}^{zz,IO}$ & $\epsilon_x^O$ &  $\epsilon_z^O$ & $\epsilon_x^I$ & $\epsilon_z^I$ & \multicolumn{2}{c}{$t_z^{TL}$} \tabularnewline 
 -0.015 & 0.012 & 0.348 & 0.152 & 0.533 & 0.410 & \multicolumn{2}{c}{$1.12\times10^{-4}$} \tabularnewline 
\hline \hline    
\end{tabular*}
\label{tab:tb}
\par\end{centering}
\end{table}

Figure \ref{fig3} displays the band structure and FS of the renormalized TL two-orbital model, with parameters provided in Table \ref{tab:tb}. 
The electron filling for the TL subsystem is 3.9, evidencing a hole doping level of 0.1 electrons relative to bulk La$_4$Ni$_3$O$_{10}$.
Notably, this actual charge-transfer direction can only be captured when the correlation effects are properly treated, as noted in \cite{39}.
The renormalized TL two-orbital model encompasses four electron pockets ($\alpha$, $\beta$, $\beta^\prime$, and $\varepsilon$) and one hole pocket ($\gamma$). 
After incorporating renormalization effects, the quasi-particle Hamiltonian yields a bandwidth of 2.1 eV. 
 The interlayer $d_{z^2}$ orbital hopping $t_{[0,0]}^{z,IO}$ is reduced from its non-interacting value of -0.650 to -0.223, which results in the flatter bonding and non-bonding bands being closer to the Fermi level. Additionally, our renormalized model reveals an intertrilayer $d_{z^2}$ orbital hopping of 0.112 meV, which is two orders of magnitude smaller than that of bulk La$_4$Ni$_3$O$_{10}$ and comparable to the interbilayer $d_{z^2}$ orbital hopping in hybrid 1212 La$_5$Ni$_3$O$_{11}$~\cite{31}.

\label{sec:SC}

\begin{figure}[t]
\centering
\includegraphics[width=0.9\columnwidth]{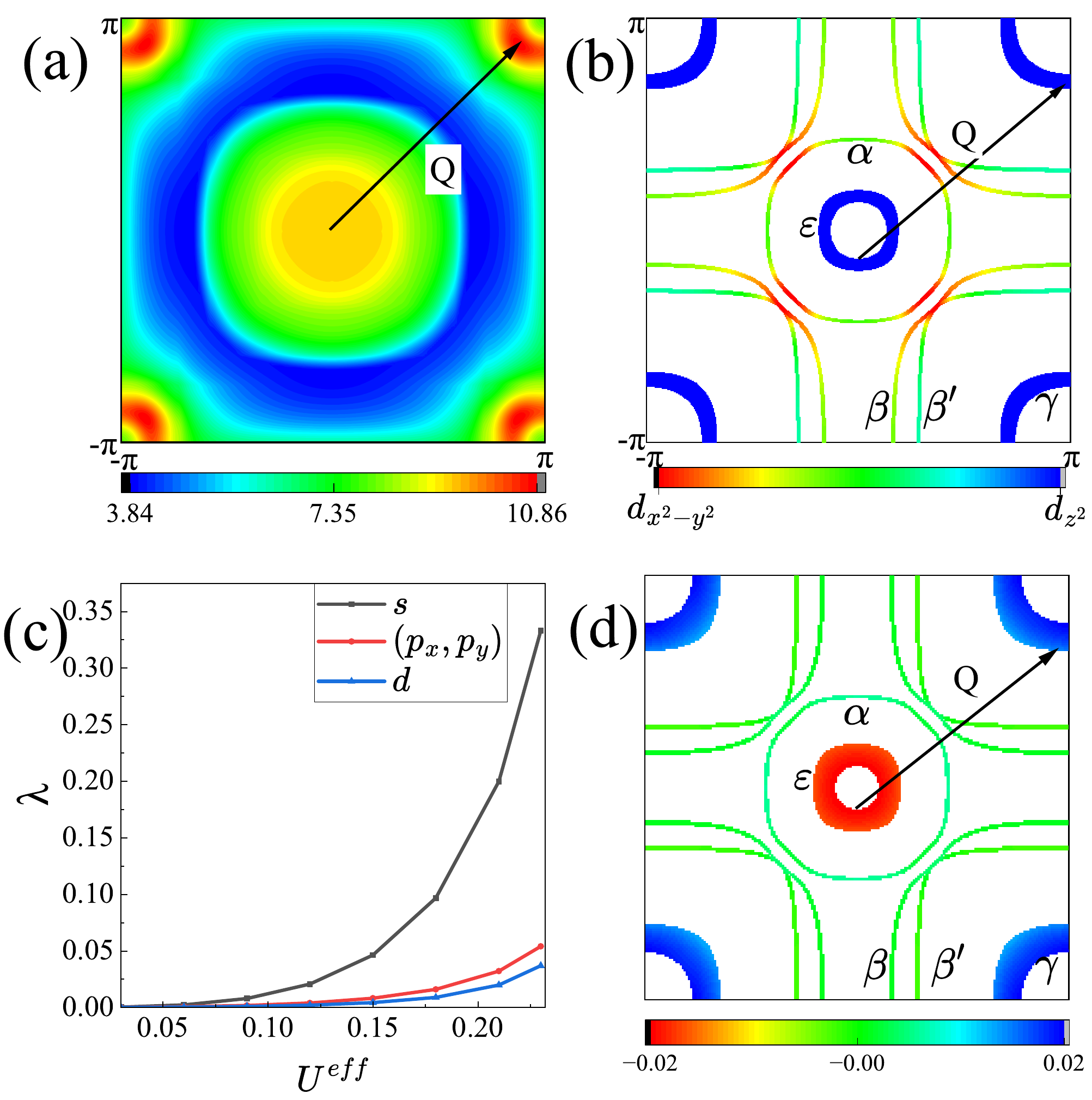}
\caption{(a) Distribution of the largest eigenvalue of the spin susceptibility matrix $\chi^{s}(q)$ in the BZ in the TL subsystem for $U^{eff} = 0.23$ eV and $J_H = U^{eff}/4$. The susceptibility peaks at Q. (b) FS of the TL subsystem in the BZ at 20 GPa. (c) The largest pairing eigenvalue $\lambda$ of the various
pairing symmetries as function of the interaction strength $U^{eff}$ with
fixed $J_H= U^{eff}/4$. (d) The distributions of the leading $s$-wave pairing gap function on the FS for $U^{eff} = 0.23$ eV.}
\label{fig4}
\end{figure}

{\it RPA study of the SC.} Since the low-energy degrees of freedom of the system are dominated by the TL subsystem, whose electronic correlations are relatively moderate, we employ the multi-orbital RPA approach (see Supplementary Material~\cite{41}) to study its SC. Based on the renormalized TB model in Eq.~\ref{eq1}, we incorporate the following Hubbard interactions:
\begin{align}\label{hubbard}
H_{int}&=U^{eff}\sum_{i\tilde{\mu}}n_{i\tilde{\mu}\uparrow}n_{i\tilde{\mu}\downarrow}+
V^{eff}\sum_{i,\sigma,\sigma^{\prime}}n_{i1\sigma}n_{i2\sigma^{\prime}} \nonumber\\
&+J_{H}^{eff}\sum_{i\sigma\sigma^{\prime}} \Big[c^{\dagger}_{i1\sigma}c^{\dagger}_{i2\sigma^{\prime}}c_{i1\sigma^{\prime}}c_{i2\sigma}+(c^{\dagger}_{i1\uparrow}c^{\dagger}_{i1\downarrow}c_{i2\downarrow}c_{i2\uparrow}+h.c.)\Big].
\end{align}
Here, $U^{eff}$, $V^{eff}$, and $J_H^{eff}$ denote the intra-orbital, inter-orbital Hubbard repulsion, and Hund’s coupling (including pair hopping) respectively, which satisfy the relation $U^{eff}=V^{eff}+2J_H^{eff}$. $U^{eff}$ can be regarded as the strength of certain residual interactions renormalized by a factor $Z^2$, where $Z$ is the quasi-particle weight. Since the quasi-particle weight $Z$ of the outer-layer $d_{z^2}$ orbital is quite small (about 0.23), $U^{eff}$ is significantly reduced compared to the unrenormalized TB value. $\tilde{\mu}$ and i denote orbital and site. By setting $J_H^{eff}=U^{eff}/4$, the critical value of $U^{eff}$ is found to be $U^{eff}_c=0.288$ eV within the RPA~\cite{42}.
Figure~\ref{fig4}(a) shows the RPA enhanced spin susceptibility $\chi^{s}(\mathbf{q},\omega=0)$ in the $q_z=0$ slice of the first Brillouin zone (BZ). The
dominant wave vector $\mathbf{Q}$ arises from nesting between the $\varepsilon$ and $\gamma$ Fermi pockets, as shown in Fig.~\ref{fig4}(b). When $U<U_c$, the spin fluctuations can mediate SC, with $T_c\propto \omega_De^{-1/\lambda}$~\cite{42}, where $\omega_D$ characterizes the spin-fluctuation energy scale. The pairing symmetry is determined by the corresponding eigenvector (see Supplementary Material~\cite{41}). Figure~\ref{fig4}(c) illustrates the dependence of the largest pairing eigenvalue $\lambda$ on the interaction strength $U^{eff}$ for different potential pairing symmetries. It is clear that a larger $U^{eff}$ leads to a stronger superconducting instability. The leading pairing symmetry is always an $s$-wave. The gap function of the obtained $s$-wave pairing is shown on the FS in Fig.~\ref{fig4}(d), which displays the $s^{\pm}$ pattern. Consequently, the $\varepsilon$- and $\gamma$- pockets connected by the nesting vector $\mathbf{Q}$ are distributed with the strongest pairing amplitude, with opposite gap signs. This pairing pattern is also similar to that in pressurized  bulk La$_4$Ni$_3$O$_{10}$~\cite{43}. 

\begin{figure}[t]
\centering
\includegraphics[width=1\columnwidth]{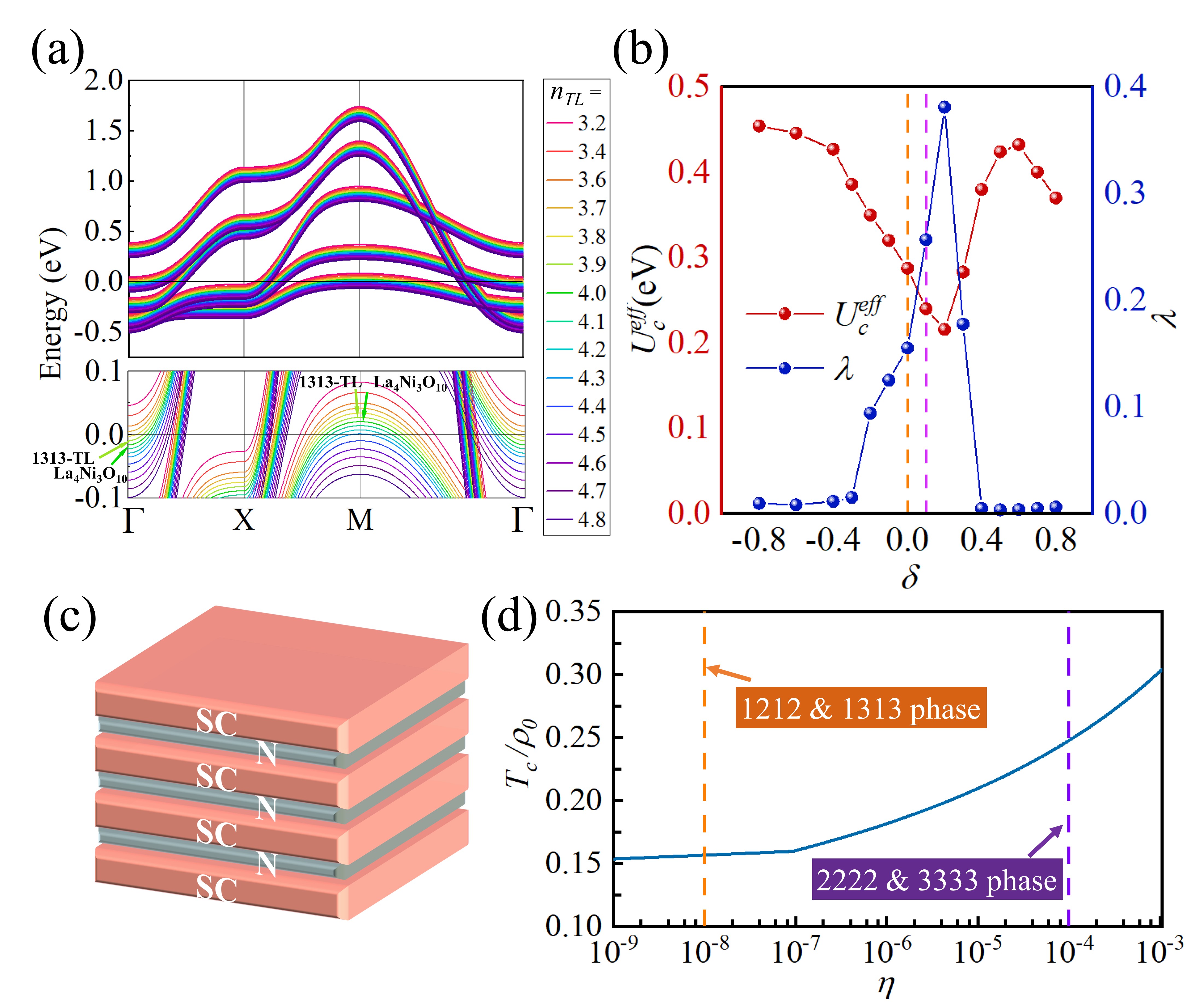}
\caption{(a) Band structures at different filling level. (b) The critical interaction strength $U_c^{eff}$ and the largest pairing eigenvalue $\lambda$ as function of doping $\delta$. The orange (purple) dashed lines mark the values of $U_c^{\mathrm{eff}}$ and $\lambda$ for the 1313 system (bulk La$_4$Ni$_3$O$_{10}$). (c) Schematic of the intrinsic S-N-S Josephson Junction structure of 1313 La$_3$Ni$_2$O$_7$, where the superconducting (S) and the normal (N) layers denote the TL and SL subsystems, respectively. (d) $T_c/\rho_0$ as a function of the anisotropy parameter $\eta$. The orange and purple dashed lines mark the $\eta$ associated with hybrid 1212 and 1313 phases, and 2222 and 3333 phases, respectively.}
\label{fig5}
\end{figure}

A natural question is why the $T_c$ in this hybrid-phase system is only 3.6 K, drastically lower than the $\sim 30$ K in bulk La$_4$Ni$_3$O$_{10}$~\cite{44}. To address this issue, we first investigate how self-doping between the SL and TL subsystems affects the electronic structure and superconducting instability in 1313 La$_3$Ni$_2$O$_7$. The band structure of the doped TL subsystem for $\delta \in (-0.8, 0.8)$ is shown in Fig.~\ref{fig5}(a). Here, $\delta = 0$ corresponds to the TL subsystem in the 1313 structure, with a filling of $n = 3.9$, slightly below $n = 4$ of bulk La$_4$Ni$_3$O$_{10}$. 
As the filling increases, the bonding and nonbonding $d_{z^2}$ bands gradually shift downward. The $\delta$ dependence of $U_c^{\mathrm{eff}}$ and $\lambda$ is shown in Fig.~\ref{fig5}(b). 
Compared to bulk La$_4$Ni$_3$O$_{10}$, the TL subsystem in the 1313 structure exhibits a larger $U_c^{\mathrm{eff}}$ but a smaller $\lambda$, indicating a lower $T_c$. To understand the underlying origin of this behavior, let us return to the FS nesting of 1313 in Fig.~\ref{fig4}(b). Around the $\Gamma$ point, the electron-like $\varepsilon$ pocket is smaller than the hole-like $\gamma$ pocket near the $M$ point. Meanwhile, in bulk La$_4$Ni$_3$O$_{10}$, the sizes of these two pockets are more comparable, indicating improved FS nesting. Better size matching enhances the nesting and thus $T_c$~\cite{43}. Overall, the reduced superconducting $T_c$ in the hybrid-phase system can be attributed to the weakened nesting condition in the TL subsystem compared to bulk La$_4$Ni$_3$O$_{10}$. Notably, recent ARPES measurements on the 1313 film reveal the disappearance of the $\gamma$ pocket, which eliminates the nesting between the $\varepsilon$ and $\gamma$ pockets and thus strongly suppresses SC~\cite{34,35}.

On the other hand, in quasi-2D layered superconductors such as 1313 La$_3$Ni$_2$O$_{7}$, the establishment of bulk SC requires not only intra-trilayer pairing but also inter-trilayer phase coherence. Given that the intermediate SL subsystem is close to a Mott-insulating state, it can be effectively regarded as a spacer layer whose itinerant electrons contribute negligibly to SC.
Physically, the inter-trilayer phase coherence is established via the coherent tunneling of a Cooper pair from one superconducting TL subsystem to an adjacent one---namely, the interlayer Josephson coupling (IJC)---across the intervening normal-metallic SL subsystems, as shown in Fig.~\ref{fig5}(c). Microscopically, this process originates from virtual single-particle tunneling events: a Cooper pair transfer involves two successive single-electron hoppings between adjacent TLs, making the IJC second order in the inter-trilayer single-particle tunneling amplitude. Furthermore, this process is extremely weak in 1313 La$_3$Ni$_2$O$_{7}$, as reflected by the very small effective TL–TL tunneling $t_z^{TL}\sim 10^{-4}$ listed in Tab.~\ref{tab:tb}.

As a consequence, the IJC leads to a stronger suppression of $T_c$ in 1313 La$_3$Ni$_2$O$_7$ than in 2222 La$_3$Ni$_2$O$_7$ and bulk La$_4$Ni$_3$O$_{10}$. It can be approximated as
\begin{align}
T_c\approx\rho_0\frac{\pi}{\mathrm{ln}(32/\eta)},
\label{eq13}
\end{align} 
where $\rho_0$ indicates the intralayer phase stiffness reflecting the pairing within the TL subsystem, which is approximated as the pairing temperature obtained from the RPA calculation, i.e. $\omega_De^{-1/\lambda}$, $\eta$ characterizes the strength of the IJC~\cite{45}. The $\eta$ for 1313 La$_3$Ni$_2$O$_7$ and bulk La$_4$Ni$_3$O$_{10}$ are compared in Fig.~\ref{fig5}(d). Since $t_z^{\mathrm{TL}} \sim 10^{-4}$ in the 1313 structure, while $t_z \sim 10^{-2}$ in bulk La$_4$Ni$_3$O$_{10}$, with $\eta$ proportional to the square of $t_z$, one obtains $\eta \sim 10^{-8}$ for the former and $\eta \sim 10^{-4}$ for the latter. As a result, the bulk $T_c$ of 1313 is lower than that of bulk La$_4$Ni$_3$O$_{10}$.  A similar IJC mechanism is also expected in other hybrid-phase systems, such as 1212, leading to a significantly lower $T_c$ compared to bulk La$_3$Ni$_2$O$_{7}$~\cite{31}.

\label{sec:conclusion}

{\it Conclusion.} We present a comprehensive investigation of the electronic and superconducting properties in the hybrid 1313 La$_3$Ni$_2$O$_7$. 
DFT+DMFT calculations reveal that the SL subsystem behaves as a nearly insulating bad metal lacking coherent quasi-particles. Therefore, SC is primarily confined within the TL subsystem. 
RPA analysis suggests that SC in 1313 La$_3$Ni$_2$O$_{7}$ exhibits an $s^{\pm}$-wave pairing symmetry. Due to the hole doping and the weakened IJC, the $T_c$ of 1313 La$_3$Ni$_2$O$_{7}$ is significantly lower than that of bulk La$_4$Ni$_3$O$_{10}$ and 2222 La$_3$Ni$_2$O$_{7}$. 
Our results demonstrate that the SL subsystem is detrimental to high-$T_c$ SC in 1313 La$_3$Ni$_2$O$_{7}$, both electronically (through Mott physics) and functionally (by weakening IJC and global phase coherence). This stands in stark contrast to the 2222 La$_3$Ni$_2$O$_{7}$, which supports robust pairing interactions and strong interlayer coherence. 
Our work establishes design rules for achieving enhanced phase coherence and high-$T_c$ in hybrid nickelate superconductors and other layered transition-metal oxides.

{\it Acknowledgements.} We are grateful for the helpful discussions with Meng Wang, Guan-Hao Feng and Zhihui Luo. This work is supported by National Natural Science Foundation of China (Grants No. 12494591, No. 92165204, No. 12574141, No. 12234016), National Key Research and Development Program of China (2022YFA1402802), Guangdong Provincial Key Laboratory of Magnetoelectric Physics and Devices (2022B1212010008), Research Center for Magnetoelectric Physics of Guangdong Province (2024B0303390001), and Guangdong Provincial Quantum Science Strategic Initiative (GDZX2401010). Ming Zhang is supported by the Zhejiang Provincial Natural Science Foundation of China under (Grant No. ZCLQN25A0402). 


\end{document}